\documentclass[journal=jacsat,manuscript=article]{achemso}
\usepackage[version=3]{mhchem} 
\usepackage[T1]{fontenc}       
\usepackage{amssymb}

\usepackage{multirow}
\usepackage{booktabs}
\usepackage{subcaption}

\author{Chao Zheng}
\email{zhengc8@mcmaster.ca}
\phone{+1647-936-1436}
\author{Oleg Rubel}
\email{rubelo@mcmaster.ca}
\phone{+1905-525-9140, ext. 24094}
\affiliation[McMaster University]
{Department of Materials Science and Engineering, McMaster University, 1280 Main Street West, Hamilton, Ontario L8S 4L8, Canada}

\title[An \textsf{achemso} demo]
  {Aziridinium lead iodide: a stable, low bandgap hybrid halide perovskite for photovoltaics}
\abbreviations{IR,NMR,UV}

\begin{document}

\begin{tocentry}

\includegraphics{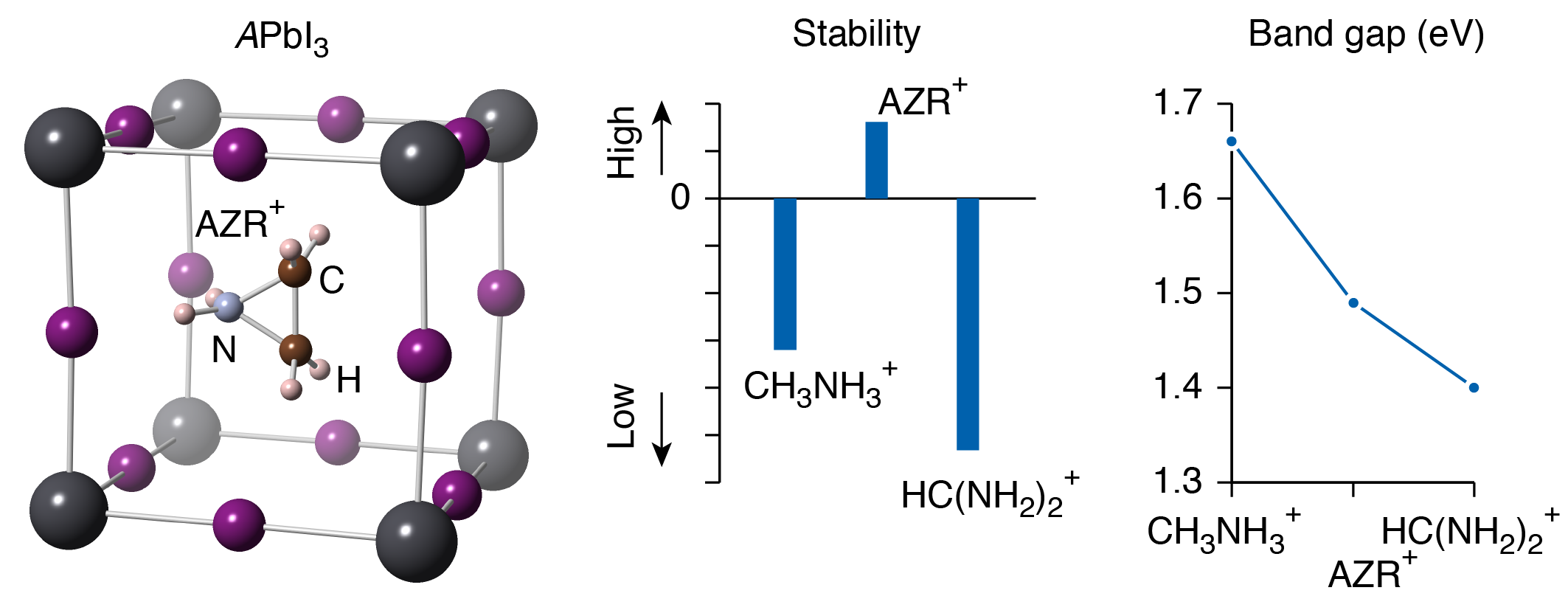}

\end{tocentry}

\begin{abstract}
The low ionization energy of an $A$ site molecule is a very important factor, which determines the thermodynamical stability of \ce{$A$PbI3} hybrid halide perovskites, while the size of the molecule governs the stable phase at room temperature and, eventually, the bandgap. It is challenging to achieve both a low ionization energy and the reasonable size for the \ce{PbI3} cage to circumvent the stability issue inherent to hybrid halide perovskites. Here we propose a new three-membered charged ring radical, which demonstrates a low ionization energy that renders a good stability for its corresponding perovskite and a reasonable cation size that translates into a suitable bandgap for the photovoltaic application. We use \textit{ab initio} calculations to evaluate a polymorphism of the crystal structure of the proposed halide hybrid perovskite, its stability and electronic properties in comparison to the mainstream perovskites, such as the methylammonium and formamidinium lead iodide. Our results highlight the importance of van der Waals interactions for predicting a correct polymorphism of the perovskite \textit{vs} hexagonal crystal structure.
\end{abstract}

\section{Introduction}
During the past ten years, halide hybrid perovskites increasingly catch researchers attentions as the absorber layers in photovoltaics \cite{Kojima_JACS_131_2009, Kim_SR_2_2012, Lee_Science_338_2012, Eperon_AFM_24_2014, Yang_Science_356_2017}. Favorable electronic properties and a low-cost fabrication method give halide hybrid perovskites an advantage over the traditional silicon. One drawback of halide hybrid perovskites is their instability. The halide hybrid perovskites easily decompose under the influence of high temperature, oxygen, water, and even UV light \cite{Christians_JACS_137_2015, Wozny_CM_27_2015, Buin_CM_27_2015}. \citet{Zhang_arXiv_2015} pointed out that the instability of methylammonium (MA) lead iodide is intrinsic due to the similar total energies of the reactant and products obtained from the density-functional theory (DFT) calculation. To commercialize the hybrid halide perovskite photovoltaics, the stability issue should be resolved.

Many different ways are proposed to stabilize the perovskite structure. Some methods involved engineering of suitable device architectures, such as changing from a liquid electrolyte to a solid hole transport material. For the representative MA lead iodide \ce{CH3NH3PbI3} based solar cells, the lifetime of hybrid halide perovskites increases from 10 minutes \cite{Im_N_3_2011} up to 1000 hours \cite{Lee_Science_338_2012} by substituting the iodide/iodine-based redox electrolyte with solid-state hole transport materials spiro-OMeTAD. Device encapsulation is also an effective strategy to prevent invasion of UV radiation, moisture or oxygen species \cite{Gevorgyan_SEMSC_116_2013, Zhou_Science_345_2014}. Although these methods indeed increase the stability of hybrid halide perovskite-based solar cells, the power conversion efficiencies still decrease by 60~\% of the initial value after 1100 hours\cite{Chauhan_JPD_2017} which is far from silicon solar panels that come on today's market with a 25-year long performance warranty. In this paper, we propose a new perovskite material with a highly unfavorable decomposition reaction enthalpy, which should stabilize it against degradation.

Hybrid halide perovskite structures discussed here are considered in the form of \ce{$A$PbI3}. Here $A$ stands for an organic radical in the lead iodide framework. Based on our recent paper, the ionization energy of the molecule on $A$ site (in addition to its size) can be an important factor which determines the stability of  perovskites \cite{Zheng_JPCC_131_2017}. A lower ionization energy of the cation favors a more stable structure. Organic molecules with the ionization energy lower than that of MA are typically much larger in size than MA. For example, the guanidinium radical \ce{C(NH2)3} and tetramethylammonium radical \ce{(CH3)4N} show low ionization energies. The radii of these cations are 278 and 320~pm, respectively, vs 217~pm for MA that precludes formation of the perovskite structure in the lead iodide framework \cite{Zheng_JPCC_131_2017}.

\begin{figure}[t]
  \includegraphics[width=0.6\textwidth]{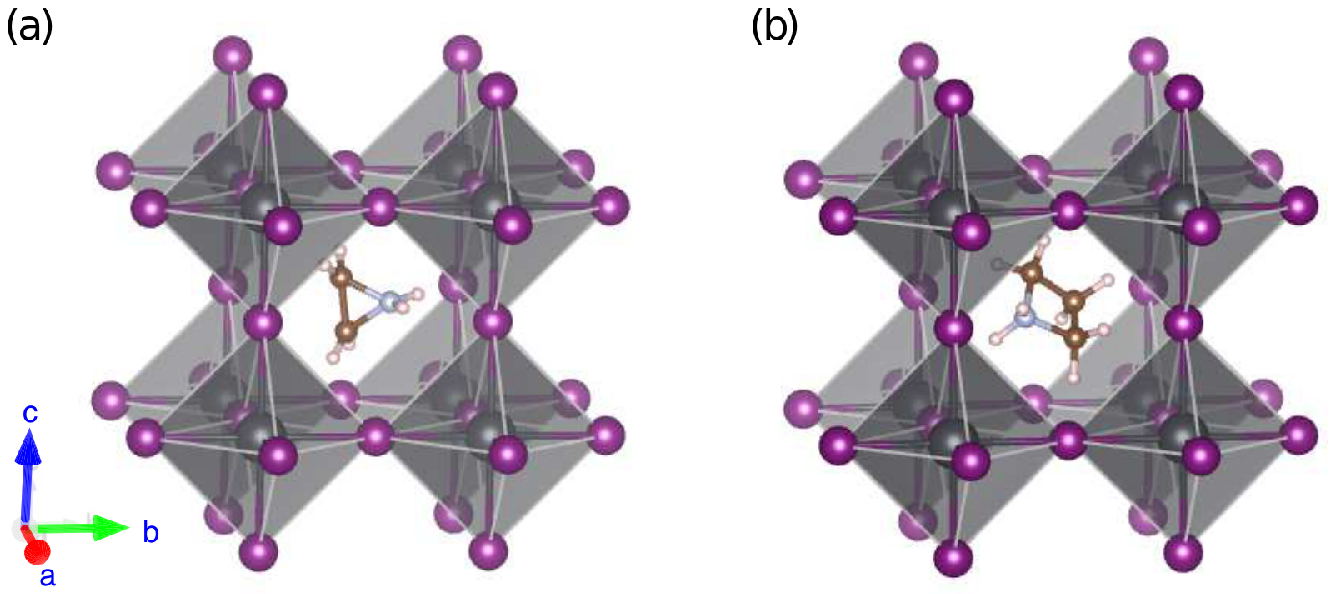}
  \caption{The quasicubic crystal structures of cyclic ring-based perovskites: (a) aziridinium lead iodide\ce{(CH2)2NH2PbI3} and (b) azetidinium lead iodide \ce{(CH2)3NH2PbI3}.}
  \label{fgr:azrpi-aztpi}
\end{figure}

\citet{Kieslich_CS_5_2014} proposed nitrogen-based cations, which were not used in perovskites before. From this group, an azetidinium radical \ce{(CH2)3NH2} is promising due to its compact structure. This four-membered ring cation is larger than MA radical, but slightly smaller than the formamidinium radical \ce{HC(NH2)2}. Recently, the azetidinium lead iodide \ce{(CH2)3NH2PbI3} was successfully synthesized \cite{Pering_JMCA_5_2017}. The crystal structure of quasicubic \ce{(CH2)3NH2PbI3} is shown in Fig.~\ref{fgr:azrpi-aztpi}b. \citet{Pering_JMCA_5_2017} reported that \ce{(CH2)3NH2PbI3} demonstrates a very good stability when soaked in water in contrast to the MA lead iodide \ce{CH3NH3PbI3}. This observation correlates with the ionization energy of the \ce{(CH2)3NH2} radical being 0.4~eV below that of MA.

Interestingly, there are also three-membered rings \cite{Solka_JPC_78_1974}, of which the aziridinium radical \ce{(CH2)2NH2} is a promising candidate to be used as the organic cation at $A$ site of hybrid halide perovskites. Figure~\ref{fgr:azrpi-aztpi}a illustrates a proposed quasicubic phase of \ce{(CH2)2NH2PbI3}. This three-membered ring cation is only slightly larger than MA. As we will show below, the ionization energy of \ce{(CH2)2NH2} is also much lower than that of MA. This fact implies that aziridinium lead iodide \ce{(CH2)2NH2PbI3} might be stable and suitable for photovoltaic applications. Here we discuss the structural stability and electronic properties of the new perovskite \ce{(CH2)2NH2PbI3}. We also compare relevant electronic properties between \ce{(CH2)2NH2PbI3}, \ce{CH3NH3PbI3}, and \ce{(CH2)3NH2PbI3}.

\section{Computational methods}

\begin{figure}[t]
  \includegraphics[width=0.4\textwidth]{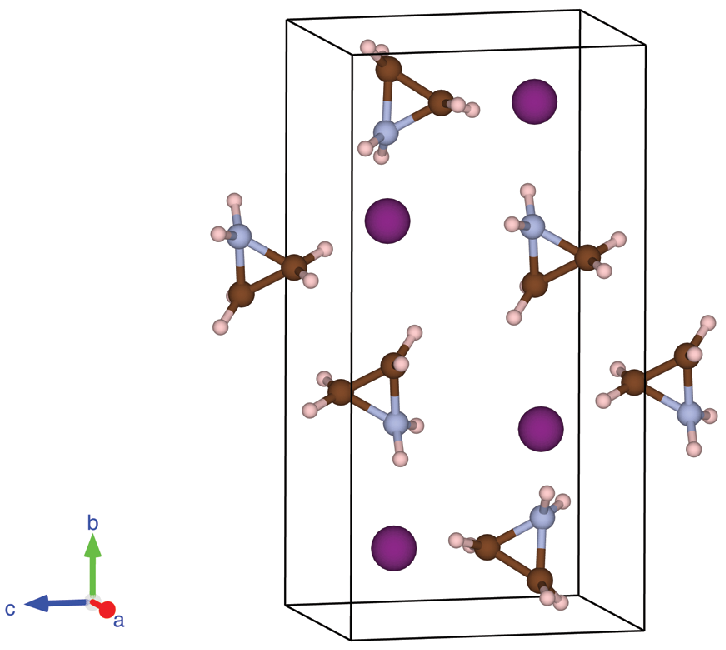}
  \caption{The crystal structure of aziridinium iodide \ce{(CH2)2NH2I}.}
  \label{fgr:azri}
\end{figure}

An experimental structure of formamidinium iodide \ce{HC(NH2)2I} \cite{Petrov_ACSECC_73_2017} was used as a parent structure for  \ce{(CH2)2NH2I} (Fig.~\ref{fgr:azri}) and \ce{(CH2)3NH2I} salts. The hexagonal structures of \ce{CH3NH3PbI3}, \ce{(CH2)2NH2PbI3} and \ce{(CH2)3NH2PbI3} are adapted from the hexagonal \ce{HC(NH2)2PbI3} \cite{Stoumpos_IC_52_2013}. The crystal structures of \ce{CH3NH3PbBr3} and \ce{CH3NH3PbCl3} are obtained from Refs.~\citenum{Swainson_JSSC_176_2003} and \citenum{Chi_JSSC_187_2005}, respectively. The crystal structures of MA halides are obtained from Refs.~\citenum{Ishida_ZNA_50_1995,Gabe_AC_14_1961,Hughes_JACS_68_1946}. Crystallographic information files (CIF) with atomic structures used in calculations can be accessed through the Cambridge crystallographic data centre (CCDC deposition numbers 1584308$-$1584327).

All calculations are based on DFT method \cite{Kohn_PR_140_1965}. Perdew-Burke-Ernzerhof (PBE) generalized gradient approximation \cite{Perdew_PRL_77_1996}  is used for the exchange-correlation functional. Dispersion interactions are included at the DFT-D3 level~\cite{Grimme_JCP_132_2010}. To predict the bandgaps of  perovskites, a many body perturbation theory, within the $GW$ approach \cite{Hedin_PR_139_1965, Hybertsen_PRB_34_1986}, is utilized including the spin-orbit coupling (SOC).  Vienna \textit{ab initio} simulation program (VASP) and projector augmented-wave potentials are used to perform all the calculations~\cite{Kresse_PRB_54_1996,Kresse_PRB_59_1999,Blochl_PRB_50_1994}.
Monopole, dipole and quadrupole corrections implemented in VASP \cite{Makov_PRB_51_4014, Neugebauer_PRB_46_16067} were used to eliminate leading errors and acquiring accurate total energies of all charged ions.

For reciprocal space integration, $4\times4\times4$ Monkhorst-Pack grid \cite{Monkhorst_PRB_13_1976} was used for cubic perovskite phases, $5\times5\times3$  for tetragonal perovskite phases, $4\times4\times4$ for hexagonal phases, $5\times5\times3$ for orthorhombic perovskite phases, and $6\times2\times4$ for the triclinic phases of \ce{(CH2)2NH2I} and \ce{(CH2)3NH2I}. The convergence of reaction enthalpies with respect to the k-mesh density is achieved better than 5~meV, which was tested by doubling the density for several perovskite structures and their reactants. The cutoff energy for a plane wave expansion was set at 500~eV. The lattice constant and atomic positions were optimized such that residual forces acting on atoms did not exceed 2~meV/{\AA}, and the residual hydrostatic pressure was less than 50~MPa.

\section{Results and discussion}

\subsection{Stability of aziridinium lead iodide \ce{(CH2)2NH2PbI3}}

Formability of inorganic perovskite structures can be rationalized via geometrical factors such as the Goldschmidt's tolerance factor~\cite{Goldschmidt_N_14_1926} and Pauling's octahedral factor \cite{Pauling_JACS_51_1929}. To calculate the geometrical factors of hybrid halide perovskites, the effective organic cation radii should be estimated. Following \citet{Kieslich_CS_5_2014} the organic cation radius can be expressed as 
\begin{equation}\label{Eq:Effective ionic radii}
	r_\text{eff} = r_\text{mass} + r_\text{ion},
\end{equation}
where $r_\text{mass}$ is the distance between the centre of mass of the molecule and the atom with the largest distance to the centre of mass (excluding hydrogen atoms) and $r_\text{ion}$ is the corresponding ionic radius of this atom. The effective radii of three- and four-membered ring cations are listed in Table~\ref{tbl:Geometrical factor and IE} along the side with the representative organic and inorganic cations. The size of cyclic cations is between that of MA and formamidinium.

The cation effective radii are used to evaluate the Goldschmidt's tolerance factor $t$ for \ce{$A$PbI3} perovskite structures (Table~\ref{tbl:Geometrical factor and IE}). The tolerance factor of \ce{(CH2)3NH2PbI3} is too large to form a cubic or tetragonal structure ($t>0.96$ \cite{Feng_JPCS_69_2008}). \citet{Pering_JMCA_5_2017} reported the experimental band gap of \ce{(CH2)3NH2PbI3} which is about 2.15~eV. If the structure of \ce{(CH2)3NH2PbI3} was tetragonal or cubic, one would expect its band gap to be less than that of \ce{CH3NH3PbI3} ($E_\text{g} < 1.6$~eV) due to a greater volume of the unit cell. The high experimental value of $E_\text{g}$ can be attributed to a different crystal structure of \ce{(CH2)3NH2PbI3}. For instance, theoretical calculations predict $E_\text{g}\approx2.6$~eV for \ce{CH3NH3PbI3} in a hexagonal phase \cite{Thind_CM_29_2017}. Similarly, \ce{HC(NH2)2PbI3} undergoes cubic to hexagonal phase transition at room temperature that is associated with an optical bleaching due to opening of the band gap \cite{Weller_JPCL_6_2015, Stoumpos_CGD_13_2013}. 


\begin{table}[t]
  \caption{Geometrical factors of selected perovskites and calculated ionization energies of corresponding $A$ site cations}
  \label{tbl:Geometrical factor and IE}
  \begin{tabular}{lcccc}
    \hline
    \multirow{2}{*}{Radical $A$} & \ce{$A$+} cation radius & Tolerance factor & \multicolumn{2}{c}{Ionization energy~(eV)}\\
    \cmidrule(l){4-5}
     & (pm) & for \ce{$A$PbI3} & DFT\textsuperscript{\emph{a}} & Exp.\\
    \hline\hline
    \ce{Cs}   & 181 &  0.81 & 3.85 & 3.89 \cite{Moore_IE_1970} \\
    \ce{CH3NH3}   & 215 & 0.91 & 4.36 & 4.30$\pm$0.1 \cite{Jeon_JACS_107_1985}\\
    \ce{(CH2)2NH2}  & 227 & 0.93 & 4.07 & $\cdots$ \\
    \ce{(CH2)3NH2} & 250 & 0.98 & 3.96 & $\cdots$\\
    \ce{HC(NH2)2} & 264 & 1.01 & 4.18 & $\cdots$ \\
    \hline
    \multicolumn{5}{l}{\textsuperscript{\emph{a}} The ionization energies are obtained from the PBE total energy difference} \\
    \multicolumn{5}{l}{~~~of cations and neutral radicals including the vibrational zero-point energy.}
  \end{tabular}
\end{table}

According to our earlier study\cite{Zheng_JPCC_131_2017}, the radical ionization energy at $A$ site has an effect on the final reaction enthalpy of hybrid halide perovskites. The lower ionization energy, the more stable is the structure. The ionization energies of corresponding organic cations are presented in Table.~\ref{tbl:Geometrical factor and IE}. Both radicals \ce{(CH2)2NH2} and \ce{(CH2)3NH2} have the ionization energies lower than MA. Their values are close to the ionization energy of Cs which is the lowest one in the periodic table. This result suggests that \ce{(CH2)2NH2PbI3} and \ce{(CH2)3NH2PbI3} should be more stable than \ce{CH3NH3PbI3}.

To investigate the stability of \ce{(CH2)2NH2PbI3}, we use the following decomposition reaction equation
\begin{equation}\label{Eq:Decomosition_reaction_general}
	\ce{(CH2)2NH2PbI3} \rightarrow  \ce{(CH2)2NH2I} + \ce{PbI2}
\end{equation}
and the corresponding enthalpy
\begin{equation}\label{Eq:Decomosition_reaction_enthalpy}
	\Delta H_\text{r} = E_\text{tot}[\ce{(CH2)2NH2PbI3}] -  E_\text{tot}[\ce{(CH2)2NH2I}] - E_\text{tot}[\ce{PbI2}]~. 
\end{equation}
The total energies $E_\text{tot}$ of products and the reactant are evaluated using DFT. If the total energy of the products is lower than the total energy of the reactant, the perovskite structure is deemed unstable \cite{Zhang_arXiv_2015}. This approach ignores finite temperature components of the free energy, which is of the order of $-0.1$~eV for \ce{CH3NH3PbI3} \cite{Tenuta_SR_6_2016}.

\begin{figure}[t]
  \includegraphics[width=0.6\textwidth]{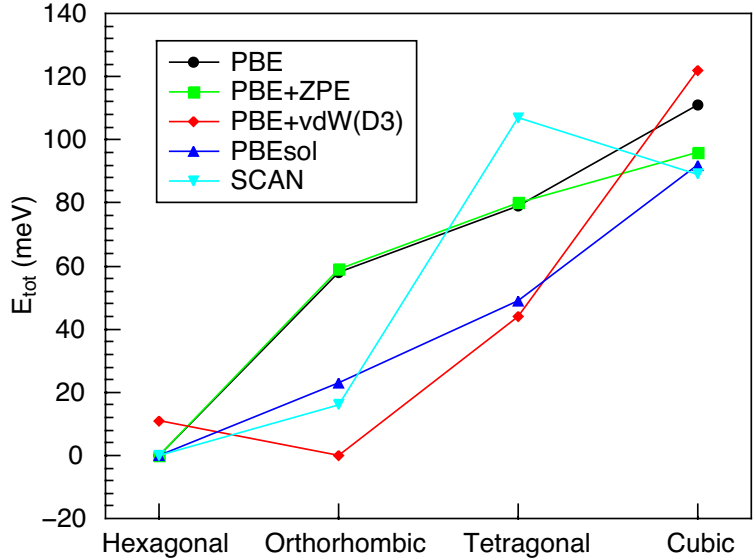}
  \caption{Polymorphism of \ce{CH3NH3PbI3} with different exchange-correlation functionals. The origin of the  energy scale is set at the lowest energy structure for each computational technique used.}
  \label{fgr:etotvspm}
\end{figure}

As a benchmark, we analyzed the stability and polymorphism of \ce{CH3NH3PbI3}. Among various polymorphs, we include a possibility for a hexagonal phase, since perovskite structures with large cations ($t>0.96$) have a tendency to adapt a hexagonal phase. PBE is first used as the exchange-correlation functional. Results listed in Table~\ref{tbl:pmofotherperovskites} suggest that the hexagonal structure of \ce{CH3NH3PbI3} has the lowest total energy at 0~K. This finding contradicts experimental data,\cite{Poglitsch_JCP_87_1987, Onoda_JPCS_51_1990, Whitfield_SR_6_2016} according to which the tetragonal \ce{CH3NH3PbI3} phase will undergo phase transformation to cubic structure above 327.4~K. The tetragonal \ce{CH3NH3PbI3} phase will transform to the orthorhombic structure below 162.2~K. During the whole polymorphism transformation of \ce{CH3NH3PbI3}, there is no hexagonal structure present. Recently, a theory-based study reported similar controversial results on the hexagonal \ce{CH3NH3PbI3} to have the lowest total energy and predicted it is the stable low-temperature phase \cite{Thind_CM_29_2017}. However, we are inclined to think that the hexagonal phase of \ce{CH3NH3PbI3} is probably a methastable phase. It is the choice of the exchange-correlation functional that is a possible reason for DFT failure to accurately predict the correct polymorphism transformation order of \ce{CH3NH3PbI3}. Thus, different exchange-correlation functionals are employed to evaluate the polymorphism transformation of \ce{CH3NH3PbI3} (see in Fig.~\ref{fgr:etotvspm}).

\begin{table}[t]
  \caption{Polymorphism of hybrid halide perovskites predicted using DFT with and without the van der Waals correction.}
  \label{tbl:pmofotherperovskites}
  \begin{tabular}{lccccc}
    \hline
    \multirow{2}{*}{Compound}  & \multirow{2}{*}{Phase} & \multicolumn{2}{c}{PBE} & \multicolumn{2}{c}{PBE+vdW(D3)}
    \\ [0.5ex]
    \cmidrule(l){3-4} \cmidrule(l){5-6}
     & & $E_\text{tot}$~(meV) & $\Delta H_\text{r}$~(meV) & $E_\text{tot}$~(meV) & $\Delta H_\text{r}$~(meV) \\
    \hline\hline
    \multirow{4}{*}{\ce{CH3NH3PbI3}}   & Cubic & 111 & 71 & 122 & 160 \\
    & Tetragonal & 79 & 39 & 44 & 82 \\
    & Orthorhombic  & 58 & 18 & 0 & 38 \\
    & Hexagonal  & 0 & -40 & 11 & 49 \\
    \hline
    \multirow{4}{*}{\ce{(CH2)2NH2PbI3}} & Cubic   & 107 & -44 & 67 & -81 \\
     & Tetragonal & 141 & -10 & 54 & -93 \\
     & Orthorhombic  & 140 & -12 & 63 & -84 \\
     & Hexagonal  & 0 & -151 & 0 & -147 \\
    \hline
    \multirow{4}{*}{\ce{(CH2)3NH2PbI3}} & Cubic   & 174 & -51 & 73 & -199  \\
     & Tetragonal & 194 & -31 & 99 & -173\\
     & Orthorhombic  & 171 & -54 & 71 & -201 \\
     & Hexagonal  & 0 & -225 & 0 & -272 \\
    \hline
    \multirow{4}{*}{\ce{HC(NH2)2PbI3}}   & Cubic & 296 & 179 & 238 & 266 \\
    & Tetragonal & 142 & 25 & 50 & 79 \\
    & Orthorhombic  & 113 & -4 & 32 & 61 \\
    & Hexagonal  & 0 & -117 & 0 & 29 \\
    \hline\hline
    \multirow{2}{*}{\ce{CH3NH3PbBr3}}   & Cubic & $\cdots$ & $\cdots$ & 71 & 13 \\
    & Orthorhombic  & $\cdots$ & $\cdots$ & 0 & -58 \\
    \hline
    \multirow{2}{*}{\ce{CH3NH3PbCl3}}   & Cubic & $\cdots$ & $\cdots$ & 68 & -41 \\
    & Orthorhombic  & $\cdots$ & $\cdots$ & 0 & -109 \\
    \hline
  \end{tabular}
\end{table}

Figure~\ref{fgr:etotvspm} illustrates the total energies of different phases of \ce{CH3NH3PbI3} obtained using PBE, PBE+ZPE (zero point energy correction), PBE+vdW(D3) \cite{Grimme_JCP_132_2010}, PBEsol \cite{Perdew_PRL_100_2008, Csonka_PRB_79_2009} and SCAN \cite{Sun_PRL_115_2015} exchange-correlation functionals. Interestingly, all methods except for the the PBE+vdW(D3) favor the hexagonal structure at low temperature. The result indicates the importance of dispersion interactions to stabilize the low-temperature orthorhombic \ce{CH3NH3PbI3}. \citet{Li_PRB_94_2016} pointed out that the van der Waals (vdW) correction is also essential to obtain an accurate lattice constant of hybrid halide perovskites. It should be noted that \citet{Thind_CM_29_2017} performed similar calculations including vdW-correction, and their results showed that the hexagonal phase still has the lowest total energy among all the phases, which contradicts our results. We suspect that the reason for discrepancy can be a suboptimal structure of the orthorhombic phase used by \citet{Thind_CM_29_2017}.

Since van der Waals effects are important for the polymorphism of \ce{CH3NH3PbI3}, the same can be extended for other types of hybrid halide perovskites. Thus we focus on results obtained with PBE+vdW(D3) exchange-correlation functional in the remaining part of the paper. Table~\ref{tbl:pmofotherperovskites} lists the results of polymorphism prediction and corresponding decomposition reaction enthalpies for the perovskites of interest here. It is noticed that except \ce{CH3NH3PbI3}, all the other perovskites favor the hexagonal structures as the low-temperature stable phase. This trend can be attributed to a greater size of organic cations involved.
For \ce{CH3NH3PbI3} and \ce{HC(NH2)2PbI3} $\Delta H_\text{r}$ values are weakly positive, which contradicts formability of those compounds. The final temperature contribution will lower the free energy down by approximately 0.1~eV \cite{Tenuta_SR_6_2016} making their formability feasible.

The Born-Haber cycles of \ce{$A$PbI3} hybrid halide perovksites indicate that the lower ionization energies of organic cations $A$ will decrease the decomposition reaction enthalpies and further stabilize the corresponding perovskites \cite{Zheng_JPCC_131_2017}. As evident from Table~\ref{tbl:Geometrical factor and IE}, the ionization energies of \ce{(CH2)2NH2} and \ce{(CH2)3NH2} are lower than those of \ce{HC(NH2)2} and MA. Accordingly, both reaction enthalpies of \ce{(CH2)2NH2PbI3} and \ce{(CH2)3NH2PbI3} in Table~\ref{tbl:pmofotherperovskites} are lower than those of \ce{CH3NH3PbI3} and \ce{HC(NH2)2PbI3}. Thus, the three- and four-memebered ring based perovskites are more stable than the currently prevalent \ce{CH3NH3PbI3} and \ce{HC(NH2)2PbI3}. 


Recent comparative studies of stability among \ce{CH3NH3Pb$X$3} perovskites with $X=$Cl, Br, and I  
reported that a higher stability can be achieved by switching halide from I, to Br and Cl \cite{Zhang_arXiv_2015, Nagabhushana_PNAS_113_2016, Jong_PRB_94_2016, Mali_NPGAM_7_2015, Maculan_JPCL_6_2015, Buin_CM_27_2015}. Our calculated stability trend of \ce{CH3NH3Pb$X$3} is consistent with those observations. The lower reaction enthalpies of \ce{(CH2)2NH2PbI3} suggest that the stability of \ce{(CH2)2NH2PbI3} will be superior to \ce{CH3NH3PbCl3} (Table~\ref{tbl:pmofotherperovskites}).

Moreover, \citet{Tenuta_SR_6_2016} indicated that the degradation of \ce{$A$PbI3} perovskites in the moist environment is governed by the solubility of a \ce{$A$I} salt. The saturation concentration $c_\text{s}$ of \ce{$A$I} in the solvent exponentially depends on the reaction enthalpy given by Eq.~(\ref{Eq:Decomosition_reaction_enthalpy}). Considering a low reaction enthalpy of \ce{(CH2)3NH2PbI3}, its decomposition via solvation of \ce{(CH2)3NH2I} in water will be hindered. This prediction is consistent with the very low solubility of \ce{(CH2)3NH2I} and the exceptional moisture stability of \ce{(CH2)3NH2PbI3} \cite{Pering_JMCA_5_2017}.

As we will see below, the band gap of \ce{(CH2)2NH2PbI3} is sensitive to its structure. For the photovoltaic application of this material, it is crucial that its structure adapts a cubic phase at the room temperature. As we discussed at the beginning of this section, the geometrical factors play an important role in formability of perovskites. \citet{Nagabhushana_PNAS_113_2016} showed the Goldschmidt's tolerance factors of \ce{CH3NH3PbI3} and \ce{CH3NH3PbBr3} are 0.91 and 0.93, respectively. The \ce{CH3NH3PbBr3} prefers cubic structure at the room temperature \cite{Poglitsch_JCP_87_1987, Onoda_JPCS_51_1990}. It is suggested that MA is a relatively small cation for the \ce{PbI3} framework, which results in a tetragonal phase at room temperature rather than the cubic one. Also, it seems that the \ce{HC(NH2)2} cation is too big for the \ce{PbI3} framework to stay as a cubic phase around room temperature. The combination of MA cation and \ce{PbBr3} framework is optimal to form a cubic phase with the tolerance factor of 0.93. Table~\ref{tbl:Geometrical factor and IE} lists the tolerance factor of 0.93 for \ce{(CH2)2NH2PbI3}. Thus, one can expect that both perovskites with exhibit a similar polymorphism as a function of temperature.

\begin{figure}[t]
  \includegraphics[width=0.6\textwidth]{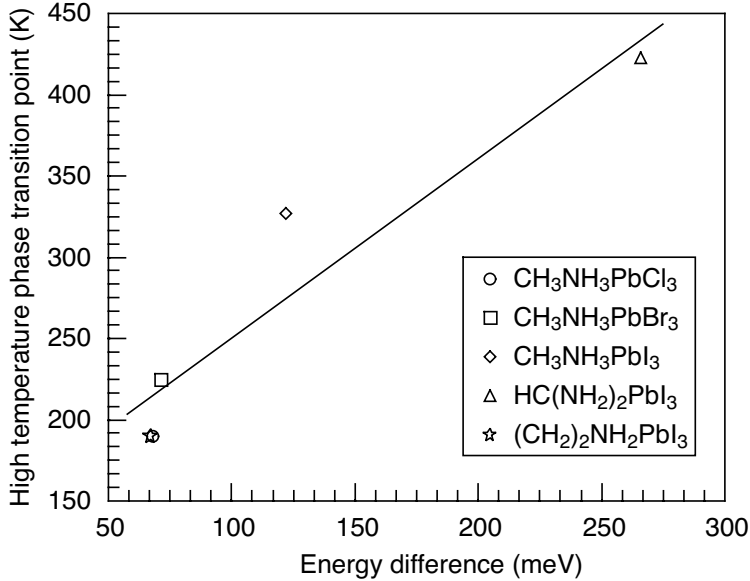}
  \caption{Correlation between the high-temperature phase transition point and the energy difference between the low-temperature phase and high-temperature phase. The linear line is a guide to the eye. High-temperature transition temperatures are taken form Refs.~\citenum{Koh_JPCC_118_2013, Poglitsch_JCP_87_1987, Onoda_JPCS_51_1990}.}
  \label{fgr:TTvsED}
\end{figure}

Besides, we found that the high-temperature phase transition point \textit{vs} the energy difference between the low-temperature  and high-temperature phases follows a linear trend (see Fig.~\ref{fgr:TTvsED}). From Fig.~\ref{fgr:TTvsED}, energy difference between cubic and hexagonal structures of \ce{(CH2)2NH2PbI3} is 67~meV, which is close to the energy difference of 68~meV between cubic and orthorhombic structures of \ce{CH3NH3PbCl3}.  The tetragonal \ce{CH3NH3PbCl3} transitions to a cubic phase at 190~K. We expect \ce{(CH2)2NH2PbI3} to have the similar transition behavior and to adapt the cubic phase above 190~K.

\subsection{Electronic properties of \ce{(CH2)2NH2PbI3}}

Till now, we found that the \ce{(CH2)2NH2} radical has a very low ionization energy and a similar Goldschmidt's tolerance factor as \ce{CH3NH3PbBr3}. Based that, the \ce{(CH2)2NH2PbI3}  perovskite solar cells should be more stable than \ce{CH3NH3PbI3} and \ce{HC(NH2)2PbI3}. The key question remained unanswered is whether the proposed perovskite structure can serve as a solar cell absorber material? Here we report the band gap for the hypothetical  \ce{(CH2)2NH2PbI3} obtained on the framework of $GW$ approximation taking into account relativistic effects. It is known that bandgap is sensitive to the structural properties. The choice of a suitable exchange-correlation functional is determined by its ability to accurately predict the lattice parameters. Since bare DFT calculations are limited to 0~K, we compare the lattice parameters calculated with different exchange-correlation functionals considering the low-temperature orthorhombic phase of \ce{CH3NH3PbI3}. Results for PBE, PBE+vdW(D3), PBEsol and SCAN functionals are listed in  Table~\ref{tbl:ex-orth-mapi}.

\begin{table}[t]
  \caption{Lattice constants prediction of orthorhombic \ce{CH3NH3PbI3} with different exchange-correlation functionals}
  \label{tbl:ex-orth-mapi}
  \begin{tabular}{lcccccccc}
    \hline
    \multirow{2}{*}{Functionals} & \multicolumn{3}{c}{Lattice constants~({\AA})} & \multicolumn{3}{c}{Error(\%)} & \multirow{2}{*}{Volume~({\AA}$^3$)} & \multirow{2}{*}{Error(\%)}
    \\ [0.5ex]
    \cmidrule(l){2-4} \cmidrule(l){5-7}
     & a & b & c & a & b & c & & \\
    \hline\hline
    Exp. \cite{Whitfield_SR_6_2016} & 8.81	& 12.59	& 8.56	& -- &	 -- &	-- &	949 & -- \\
    PBE & 9.23		& 12.86	& 8.63	& 4.8 & 2.2 &	0.8 &	1025	& 7.9 \\
    PBE+vdW(D3) & 8.92	& 12.72	& 8.51  & 1.2 & 1.1 &	$-$0.5 &	966	& 1.7 \\
    PBEsol & 8.96  & 12.61	& 8.43	& 1.7 & 0.2 &	$-$1.5 &	953 &	 0.4 \\
    SCAN & 8.93	& 12.69	& 8.53	& $-$0.4 & $-$0.3 &	$-$0.4	& 974 & 2.6\\
    \hline
  \end{tabular}
\end{table}

We found that the PBE+vdW(D3), PBEsol, and SCAN exchange-correlation functionals can provide a reasonable prediction for the lattice constants. Here, we continue to use PBE+vdW(D3) optimized perovskite structures in order to remain consistent with the section on stability calculations. Results for bandgaps obtained using PBE+vdW(D3) exchange-correlation functional with and without SOC effect are shown in Table~\ref{tbl:BGofotherperovskites}.

\begin{table}[t]
  \caption{Bandgaps (eV) of hybrid halide perovskites prediction with PBE+vdW(D3) and PBE+vdW(D3)+SOC}
  \label{tbl:BGofotherperovskites}
  \begin{tabular}{lccc}
    \hline
    Compound  & Phase & PBE+vdW(D3) & PBE+vdW(D3)+SOC \\

    \hline\hline
    \multirow{4}{*}{\ce{CH3NH3PbI3}}   & Cubic & 1.44 &  0.38 \\
    & Tetragonal & 1.51 & 0.76 \\
    & Orthorhombic  & 1.72 & 0.86 \\
    & Hexagonal  & 2.53 & 2.20 \\
    \hline
    \multirow{4}{*}{\ce{(CH2)2NH2PbI3}} & Cubic & 1.35 & 0.36 \\
     & Tetragonal & 1.58 &  0.62 \\
     & Orthorhombic  & 1.53 & 0.58  \\
     & Hexagonal  & 2.74 & 2.35 \\
    \hline
    \multirow{4}{*}{\ce{(CH2)3NH2PbI3}} & Cubic   & 1.56 & 0.54  \\
     & Tetragonal & 1.61 &  0.64 \\
     & Orthorhombic  & 1.71 &  0.67 \\
     & Hexagonal  & 2.69 & 2.31 \\
    \hline
    \multirow{4}{*}{\ce{HC(NH2)2PbI3}}   & Cubic & 1.34 &  0.33 \\
    & Tetragonal & 1.60 & 0.67 \\
    & Orthorhombic  & 1.66 & 0.67 \\
    & Hexagonal  & 2.52 & 2.20 \\
    \hline
  \end{tabular}
\end{table}

It is well known that DFT calculations with SOC grossly underestimates the bandgap of perovskites \cite{Umari_SR_4_2014}. In Table~\ref{tbl:BGofotherperovskites}, the calculated bandgaps with PBE+vdW(D3)+SOC are approximately 1~eV lower than the result without SOC. Bandgaps increase from cubic phase to hexagonal phase. Polymorphs of \ce{(CH2)3NH2PbI3} demonstrate the largest bandgaps among the four perovskites studied here. Bandgaps of different phases of \ce{(CH2)2NH2PbI3} lies between \ce{CH3NH3PbI3} and \ce{HC(NH2)2PbI3}. It is obvious that hexagonal phases show much large bandgaps than perovskite phases. Thus, the hexagonal phase around room temperature is not desirable when aiming at photovoltaic applications. 

\begin{figure}[t]
  \includegraphics[width=0.6\textwidth]{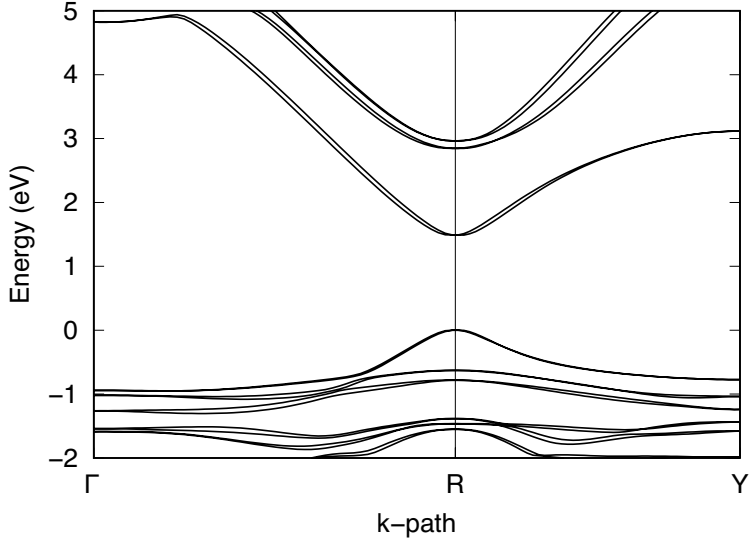}
  \caption{Electronic band structure of \ce{(CH2)2NH2PbI3} calculated at the PBE+SOC level with the band gap adjusted to match the $G_0W_0+$SOC result in Table~\ref{tbl:gw-BG}. The $k$-path includes the following high-symmetry points: $\Gamma(0, 0, 0)$, R$(0.5, 0.5, 0.5)$, and Y$(0, 0.5, 0)$.}
  \label{fgr:BSofAZRPI}
\end{figure}

Figure~\ref{fgr:BSofAZRPI} shows a relativistic band structure of quasicubic \ce{(CH2)2NH2PbI3}. The fundamental band gap is near R-point of the Brillouin zone. The presence of a Rashba splitting is noticeable at the vicinity of the band extrema, however its magnitude is heavily reduced when compared with the Rashba splitting at the \ce{CH3NH3PbI3} band edges \cite{Even_JPCL_4_2013, Brivio_PRB_89_2014, Zheng_NL_15_2015, Rubel_2015_arXiv}.  The Rashba splitting in halide hybrid perovskites originates from the strong spin-orbit interaction and distortions in the Pb-centred octahedron \cite{Zheng_NL_15_2015, Rubel_2015_arXiv}. A more centrosymmetric \ce{PbI3} cage of the quasicubic structure and its larger volume can be a reason for the reduced Rashba splitting in \ce{(CH2)2NH2PbI3}.

The Rashba splitting in hybrid halide perovskites introduces an effectively indirect bandgap, which prolongs the carrier lifetime \cite{Zheng_NL_15_2015,Stier_NC_7_2016}. We expect the reduced Rashba splitting in \ce{(CH2)2NH2PbI3} not to impede its power conversion efficiency for photovoltaic applications, since a similar weaker splitting can be found in the band structures of \ce{HC(NH2)2PbI3} \cite{Chen_PNAS_114_2017}. Recently, \citet{Yang_Science_356_2017} fabricated formamidinium-lead-halide-based perovskite solar cells with a certified 22.1\% power conversion efficiency.

It is well known that a first-principle $GW$ approximation with SOC can accurately predict the bandgaps of hybrid halide perovskites \cite{Umari_SR_4_2014, Brivio_PRB_89_2014}. First, we used a single-shot $G_0W_0$ approximation with SOC to evaluate the bandgaps of quasicubic phases for perovskites of interest (Table~\ref{tbl:gw-BG}). Next, we carried out a partially self-consistent calculation by performing 4 iterations of $G$ only (referred to as $GW_0$). The single-shot $G_0W_0$ method gave the best match of bandgaps with the experimental values (Table~\ref{tbl:gw-BG}). Based on this table, we found that the bandgap of quasicubic \ce{(CH2)2NH2PbI3} is 0.17~eV lower than quasicubic \ce{CH3NH3PbI3} and 0.09~eV higher than quasicubic \ce{HC(NH2)2PbI3}. This result suggests that \ce{(CH2)2NH2PbI3} may offer a superior utilization of the Sun's spectrum than \ce{CH3NH3PbI3}.

\begin{table}[t]
  \caption{Quasiparticle bandgaps (eV) of various perovskites in the quasicubic phase.}
  \label{tbl:gw-BG}
  \begin{tabular}{lcccc}
    \hline
    Perovskites                            & $G_0W_0+$SOC  & $GW_0+$SOC & Exp. & Reported $GW+$SOC\\
    \hline
    \ce{CH3NH3PbI3}   & 1.66   & 1.76 & 1.69 \cite{Quarti_EES_9_2016}  & 1.27$-$1.67 \cite{Brivio_PRB_89_2014, Quarti_EES_9_2016, Ahmed_EPL_108_2015, Bokdam_SR_6_2016, Gao_PRB_93_2016} \\
    \ce{(CH2)2NH2PbI3} & 1.49  & 1.53 & $\cdots$ & $\cdots$ \\
    \ce{(CH2)3NH2PbI3}   & 1.84   & 1.99 & $\cdots$ & $\cdots$ \\
    \ce{HC(NH2)2PbI3} & 1.40  & 1.44 & 1.43$-$1.48 \cite{Koh_JPCC_118_2013, Pang_CM_26_2014, Eperon_EES_7_2014, Stoumpos_ACR_48_2015} & 1.46 \cite{Arora_ACSEL_1_2016} \\
    \hline    
    
  \end{tabular}

\end{table}

\section{Conclusions}

In this paper, we proposed a three-membered cyclic organic cation based halide hybrid perovskite \ce{(CH2)2NH2PbI3} which has a potential to be used as the absorber material for photovoltaics. The low ionization energy of organic radical \ce{(CH2)2NH2} decreases the reaction enthalpy of forming the corresponding perovskite. It suggests that this lower reaction enthalpy renders a much better stability of \ce{(CH2)2NH2PbI3} than \ce{CH3NH3PbI3} and \ce{HC(NH2)2PbI3}. The appropriate cation radius of \ce{(CH2)2NH2} for the \ce{PbI3} framework and a low energy difference between high-temperature and low-temperature phases make \ce{(CH2)2NH2PbI3} transfer to a cubic phase feasible below the room temperature. Besides, we found that DFT calculation with PBE+vdW(D3) as the exchange-correlation functional can predict the correct order of polymorphism of \ce{CH3NH3PbI3}.  Relativistic band structure plot demonstrates the existence of a Rashba splitting in \ce{(CH2)2NH2PbI3}, albeit  less prominent than in \ce{CH3NH3PbI3}. The Rashba splitting will allow \ce{(CH2)2NH2PbI3} to form an indirect bandgap near R-point in the Brillouin zone and benefit from an enhanced charge carrier lifetime. $GW$ calculations suggest that the cubic phase of \ce{(CH2)2NH2PbI3} has an even lower bandgap of 1.49 eV than \ce{CH3NH3PbI3} thereby making the former perovskite a suitable absorber material for solar cells.

\begin{acknowledgement}

Funding was provided by the Natural Sciences and Engineering Research Council of Canada under the Discovery Grant Program RGPIN-2015-04518. The work was performed using a Compute Canada infrastructure.

\end{acknowledgement}

\bibliographystyle{manuscri}

\providecommand{\latin}[1]{#1}
\providecommand*\mcitethebibliography{\thebibliography}
\csname @ifundefined\endcsname{endmcitethebibliography}
  {\let\endmcitethebibliography\endthebibliography}{}

\end{document}